\documentclass[journal]{IEEEtran}

\usepackage{graphicx}
\usepackage{amsmath}
\usepackage{multirow}

\usepackage[justification=centering]{caption}

\ifCLASSINFOpdf
\else
\fi
\begin{document}
%
\title{IP-UNet: Intensity Projection UNet Architecture for 3D Medical Volume Segmentation}
%
%
%

\author{Nyothiri Aung, Tahar Kechadi, Liming Chen and Sahraoui Dhelim
\thanks{Nyothiri Aung, Tahar Kechadi and Sahraoui Dhelim are with the School of Computer Science, University College Dublin, Ireland.} 
\thanks{Liming Chen is with the School of Computing, Ulster University, U.K.}
}

%
%

\markboth{Journal of \LaTeX\ Class Files,~Vol.~14, No.~8, August~2015}%
{Shell \MakeLowercase{\textit{et al.}}: Bare Demo of IEEEtran.cls for IEEE Journals}
%



\maketitle

\begin{abstract}
CNNs have been widely applied for medical image analysis. However, limited memory capacity is one of
the most common drawbacks  of processing high-resolution 3D volumetric data.  3D volumes are usually
cropped or  downsized first before processing,  which can result  in a loss of  resolution, increase
class  imbalance, and  affect the  performance of  the segmentation  algorithms. In  this paper,  we
propose an  end-to-end deep  learning approach called  IP-UNet. IP-UNet is  a UNet-based  model that
performs multi-class segmentation on Intensity Projection (IP)  of 3D volumetric data instead of the
memory-consuming 3D volumes. IP-UNet uses limited  memory capability for training without losing the
original 3D image  resolution. We compare the  performance of three models in  terms of segmentation
accuracy and  computational cost: 1) Slice-by-slice  2D segmentation of  the CT scan images  using a
conventional 2D  UNet model.  2) IP-UNet  that operates on  data obtained  by merging  the extracted
Maximum  Intensity  Projection  (MIP),  Closest  Vessel  Projection  (CVP),  and  Average  Intensity
Projection (AvgIP)  representations of the source  3D volumes, then  applying the UNet model  on the
output IP  images. 3) 3D-UNet model  directly reads the 3D  volumes constructed from a  series of CT
scan images  and outputs the  3D volume of  the predicted segmentation.  We test the  performance of
these methods  on 3D volumetric  images for  automatic breast calcification  detection. Experimental
results show  that IP-Unet  can achieve  similar segmentation  accuracy with  3D-Unet but  with much
better performance. It reduces the training time by 70\% and memory consumption by 92\%. 

\end{abstract}


%
\IEEEpeerreviewmaketitle

\section{Introduction}
\label{sec:Intro}
Deep convolution  neural networks  (CNN) have proven  their efficiency for  image analysis,  such as
Deep convolution  neural networks  (CNN) have proven  their efficiency for  image analysis,  such as
image classification and  image segmentation \cite{cnn}. CNNs have been  widely applied successfully
for  medical  image analysis.  Particularly  UNet  \cite{unet},  a  CNN architecture,  had  achieved
satisfactory results in medical image segmentation.  Medical images such as Computed Tomography (CT)
and Magnetic Resonance Imaging (MRI) are represented as 3D volumes rather than 2D images. Therefore,
there  are   3D  variants   of  CNN   architectures,  such  as   3D-CNN  \cite{3dcnn}   and  3D-UNet
\cite{3dunet}. These 3D models usually yield better  accuracy, as they can capture the 3D-contextual
information of volumetric  data rather than individually processing CT/MRI  slices. However, limited
memory capacity  is one  of the most  common drawbacks of  processing high-resolution  3D volumetric
data. 3D  volumes are  usually cropped or  downsized first before  processing. These  operations can
result in a loss of resolution and increase the class imbalance in the input data batches, which can
affect  the performances  of segmentation  algorithms. This  problem is  known as  memory bottleneck
\cite{bottleneck}.

Intensity  projection  in medical  imaging  is  used  to transform  3D  images  into 2D  images  for
visualization purposes. For instance, we use maximum intensity projection (MIP) to project the voxel
with the  highest Hounsfield unit  on a given  axis throughout  the volume onto  a 2D image.  MIP is
effective   for   automatic   pulmonary   nodule  detection   \cite{mip4nodule},   tumor   detection
\cite{mip4tumor}, and  rheumatoid arthritis  detection \cite{mip4rheumatoid}, to  name a  few. Local
maximum intensity projection (LMIP), also known as  Closest Vessel Projection (CVP), was proposed to
improve MIP. LMIP  selects the local maximum  Hounsfield unit above a certain  threshold rather than
the absolute  maximum, as  in MIP.  Average intensity projection  (AvgMIP) \cite{avgIP}  and Minimum
intensity projection (MinIP) \cite{minip} are also widely  used in medical imagining. In this paper,
we leverage  intensity projection to  perform segmentation on the  projection of 3D  volumes without
loss of  precision introduced  by cropping  or downsizing.  Our contributions  can be  summarized as
follows: 

\begin{itemize}
    \item  We  propose  IP-UNet, an  end-to-end  deep  learning  model  that operates  on  intensity
      projections of 3D volumes  instead of the memory-consuming 3D volumetric  data. IP-UNet can be
      trained with limited memory capability without loss of resolution of the original 3D images. 
    \item We compare the performance of IP-UNet  in terms of segmentation accuracy and computational
      cost with similar baselines, namely, UNet and 3D-UNet. 
    \item  We  train  and  evaluate  IP-UNet  on 500  3D  volumetric  images  for  automatic  breast
      calcification detection. 
\end{itemize}

The  rest  of  the  paper  is  organized   as  follows:  Section  \ref{sec.2}  reviews  the  related
literature. In  Section \ref{sec.3}, we present  the proposed IP-UNet architecture  architecture. In
Section  \ref{sec.4}, we  compare  IP-UNet to  similar models  regarding  segmentation accuracy  and
computational cost. Finally, Section \ref{sec.5} concludes the paper.

\section{Related work}
\label{sec.2}
Many previous works have studied computational and memory efficiency when dealing with 3D volumetric data. 

Kozinski et al. \cite{Kozinski2020,Kozinski2018} introduced a loos function for deep neural networks that can be used for 3D volumetric data segmentation. They proposed annotating 2D projections of the training 3D volumes, which reduces the amount of effort needed to annotate the volumes. Their results show that only a marginal accuracy loss is introduced by their method, by in return they get a considerable reduction in annotation effort. Wang et al. \cite{Wang2018} proposed a two-stage Unet framework that simultaneously learns to identify a Region of Interest (ROI) within the entire 3D volume and to classify voxels without losing the original resolution. Their proposed method classifies all the voxels viewed from an axial slice based on a pre-defined neighborhood of axial slices around it.

Similarly, Zheng et al \cite{zheng2020deep,zheng2019automatic} trained separately various deep-learning models using the MIP of CT images with different slab thicknesses. They concluded that for a single slab thickness, MIP images of 10 mm are the most effective for lung nodule detection. Specifically, The sensitivity increased (82.8\% to 90.0\%) for slab thickness of 1 to 10 mm and decreased (88.7\% to 76.6\%) for slab thickness of 15–50 mm. Their results proved that the use of images with multiple MIP settings can improve the performance at the nodule candidate detection stage. Furthermore, the optimal MIP slab thickness in their deep-learning model was similar to the one commonly used by radiologists in the clinic. Fujioka et al. \cite{fujioka2021deep,takahashi2022deep} compared the performance of various CNN models (DenseNet121, DenseNet169, InceptionResNetV2, InceptionV3, NasNetMobile, and Xception) when applied to MIP of dynamic contrast-enhanced breast MRI images. They find out that operating CNN on MIP can achieve up to 98\% AUC. Wang et al \cite{wang2022artificial} applied ResNet-50 with the axial and sagittal MIP of early post-contrast subtracted breast MRI images. Their system achieved comparable diagnostic performance with the senior radiologist and outperformed the junior radiologist. Song et al \cite{song2022} proposed a 3d-UNet segmentation network for coronary artery detection, which is able to extract and fuse rectifying features efficiently. They used dense blocks in 3D-UNet for the encoding process to obtain rich representative features related to coronary artery segmentation; and 3D residual blocks with feature rectification capability are used to improve the segmentation accuracy. Similarly, Lei et al \cite{lei2020} introduced a 3D attention fully convolution network (FCN) method for automatic segmentation of the coronary artery. FCN was employed to operate end-to-end mapping from coronary computed tomographic angiography images to the binary segmentation of coronary artery. Deep attention strategy was used to highlight the informative semantic features extracted from CCTA image and to improve the segmentation accuracy.

\begin{figure*}[!htb]
	\centering
	\includegraphics[scale=0.5]{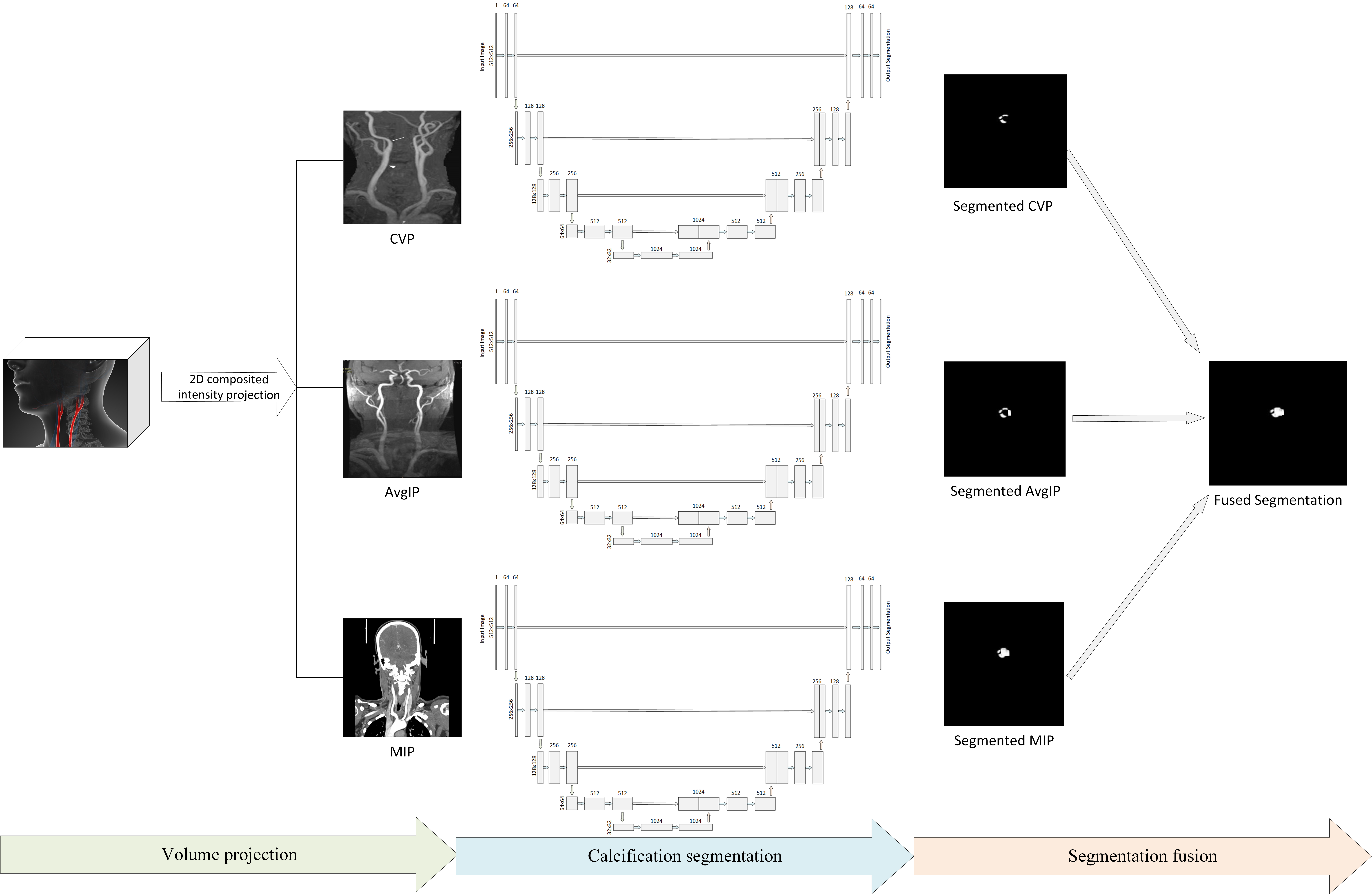}
	\caption{IP-UNet overall architecture}
	\label{ipunet}
\end{figure*}

\begin{figure*}[!htb]
	\centering
	\includegraphics[scale=0.9]{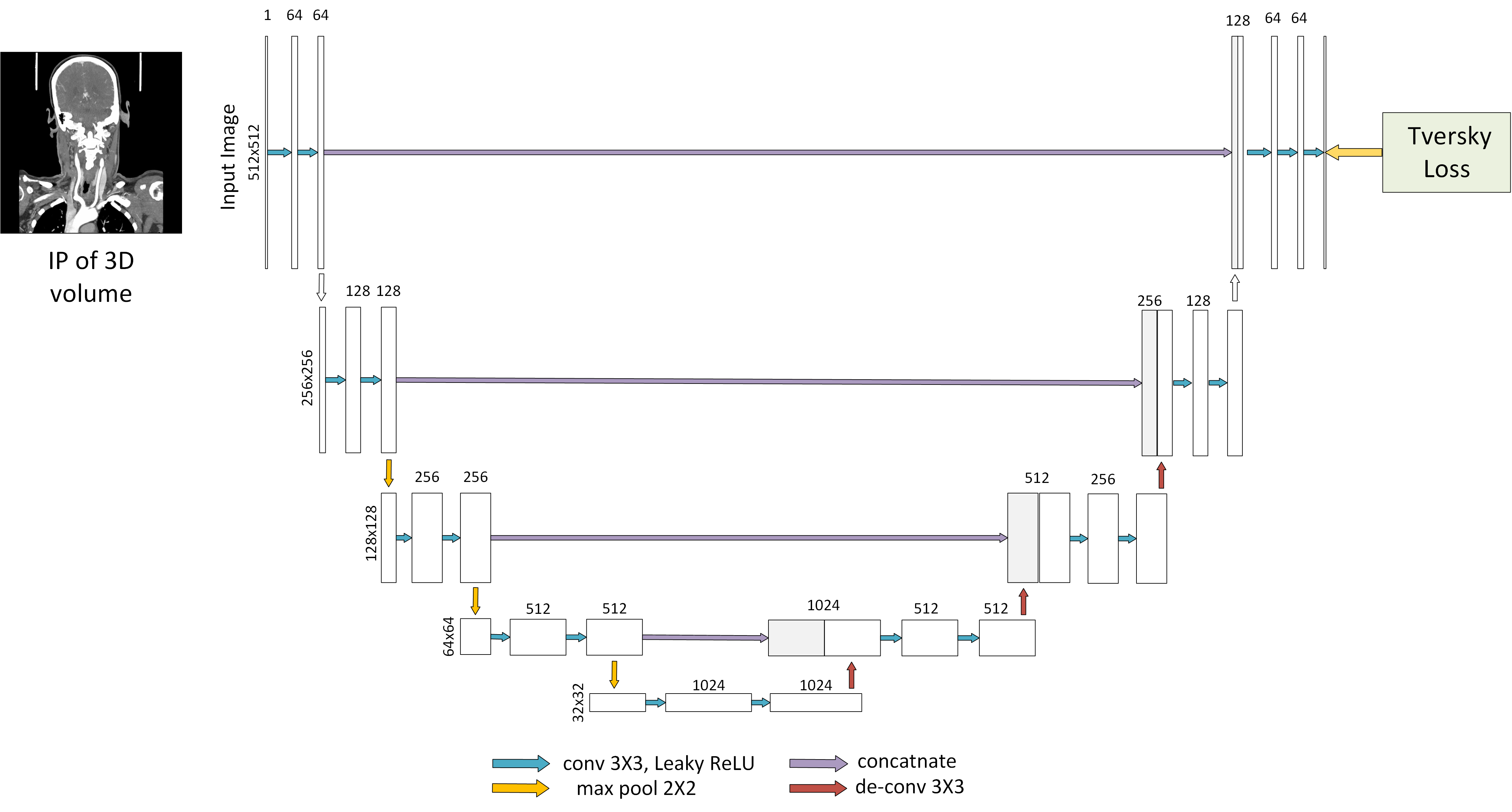}
	\caption{IP-UNet layers architecture}
	\label{unet}
\end{figure*}

\section{IP-UNet architecture}
\label{sec.3}

The general architecture of IP-UNet is shown in Figure \ref{ipunet}. IP-UNet takes the 3D volumetric image as input, and applies three intensity projection functions on the sagittal axis of the 3D volume, namely LMIP, AvgIP and MIP, as shown in eq(\ref{cvp_eq}), eq(\ref{avgip_eq}) and eq(\ref{mip_eq}) respectively. Where, $v_{i,j}$ is the projected voxel in the new 2D image in position $(i,j)$, and $v_{i,j,k}$ is the voxel the 3D image in position $(i,j,k)$, and $x$ is the local maximum intensity threshold for CVP. It is worth noting that we applied intensity projection functions of the sagittal axis because coronary artery calcification is salient from the sagittal view. Intensity projection functions can be also applied on the coronal axis or axial axis depending on the use case. For instance, in the context of brain image analysis, applying intensity projection function on the axial axis could be most applicable.  Figure \ref{unet} shows the neural network architecture IP-UNet. The encoder and decoder phases follow the original U-Net model. Each encoding level includes two 3 × 3 convolutions operators, each operator is followed by a batch normalization and a Leaky ReLU operation. We leverage LeakyReLU \cite{dubey2019}, instead of ReLU function used in the original UNet, to mitigate the dying Relu problem, where most of the neural neurons are set to zero because of negative inputs during the learning process.  For down-sampling, each encoding level is followed by a 2×2 max pooling operation with the stride of two, Followed by convolution operations that double the feature channel count as detailed in Table \ref{unet_parameters}. In each decoding level, the height and width of the feature maps is doubled through up-sampling. Then the resulting block is concatenated with the corresponding feature maps block from the encoding phase.

\begin{equation}
\label{cvp_eq}
{
CVP(v_{i,j})= 
\begin{cases}
   max([v_{i,j,0},...,v_{i,j,k-1}]),& \text{if} \quad max \leq x\\
    0,              & Otherwise
\end{cases}}
\end{equation}

\begin{equation}
\label{avgip_eq}
{AvgIP(v_{i,j})= \frac{\sum{[v_{i,j,0},...,v_{i,j,k-1}]}}{\|[v_{i,j,0},...,v_{i,j,k-1}]\|}}
\end{equation}

\begin{equation}
\label{mip_eq}
{MIP(v_{i,j})= max([v_{i,j,0},...,v_{i,j,k-1}])}
\end{equation}
IP-UNet applies 2D UNet architecture on each of the composited 2D intensity projections. In our case, the input images are 512X512. Each of the 2D projections passes through encoding and decoding phases. In the encoding phase, each level includes 3X3 convolutions, followed by batch normalization and ReLU. For each downsampling step, 2X2 max pooling with a stride of 2 is applied. As shown in Table \ref{unet_parameters}, there are 14 layers in the encoding phase, 10 convolutional layers, plus 4 max pooling layers. In the decoding phase, the size of feature maps is doubled step through de-convolution and concatenation with copies from the decoding phase, which allows the fusing of shallow high-resolution features with deep semantic features. Table \ref{unet_parameters} shows the layers of encoding and decoding phases. 

\begin{table}
\caption{IP-UNet network layers}
\label{unet_parameters}
\centering
\begin{tabular}{|p{0.15cm}|l|p{0.6cm}|l|p{0.5cm}|p{0.6cm}|l|} 
\hline
\textbf{No} & \textbf{Phase} & \textbf{Type} & \textbf{Input} & \textbf{Filter} & \textbf{Stride/ Size} & \textbf{Output}  \\ 
\hline
1           & Encode         & conv          & 512X512X2      & 64               & 1/3X3                & 512X512X64       \\ 
\hline
2           & Encode         & conv          & 512X512X64     & 64               & 1/3X3                & 512X512X64       \\ 
\hline
3           & Encode         & max           & 512X512X64     &                  & 2/2X2                & 256X256X64       \\ 
\hline
4           & Encode         & conv          & 256X256X64     & 128              & 1/3X3                & 256X256X128      \\ 
\hline
5           & Encode         & conv          & 256X256X128    & 128              & 1/3X3                & 256X256X128      \\ 
\hline
6           & Encode         & max           & 256X256X128    &                  & 2/2X2                & 128X128X128      \\ 
\hline
7           & Encode         & conv          & 128X128X128    & 256              & 1/3X3                & 128X128X256      \\ 
\hline
8           & Encode         & conv          & 128X128X256    & 256              & 1/3X3                & 128X128X256      \\ 
\hline
9           & Encode         & max           & 64X64X256      &                  & 2/2X2                & 64X64X256        \\ 
\hline
10          & Encode         & conv          & 64X64X512      & 512              & 1/3X3                & 64X64X512        \\ 
\hline
11          & Encode         & conv          & 64X64X512      & 512              & 1/3X3                & 64X64X512        \\ 
\hline
12          & Encode         & max           & 32X32X512      &                  & 2/2X2                & 32X32X512        \\ 
\hline
13          & Encode         & conv          & 32X32X1024     & 1024             & 1/3X3                & 32X32X1024       \\ 
\hline
14          & Encode         & conv          & 32X32X1024     & 1024             & 1/3X3                & 32X32X1024       \\ 
\hline
15 & Decode & deconv & 32X32X1024 & 512 & 2/3X3 & 64X64X512 \\
\hline
16 & Decode & concat & 64X64X512 &  &  & 64X64X1024 \\
\hline
17 & Decode & conv & 64X64X1024 & 512 & 1/3X3 & 64X64X512 \\
\hline
18 & Decode & conv & 64X64X512 & 512 & 1/3X3 & 64X64X512 \\

\hline
19 & Decode & deconv & 64X64X512 & 256 & 2/3X3 & 128X128X256 \\
\hline
20 & Decode & concat & 128X128X512 &  &  & 128X128X512 \\
\hline
21 & Decode & conv & 128X128X512 & 256 & 1/3X3 & 128X128X256 \\
\hline
22 & Decode & conv & 128X128X256 & 256 & 1/3X3 & 128X128X256 \\
\hline
23 & Decode & deconv & 128X128X256 & 128 & 2/3X3 & 256X256X128 \\
\hline
24 & Decode & concat & 256X256X128 &  &  & 256X256X256 \\
\hline
25 & Decode & conv & 256X256X256 & 128 & 1/3X3 & 256X256X128 \\
\hline
26 & Decode & conv & 256X256X128 & 128 & 1/3X3 & 256X256X128 \\

\hline
27 & Decode & deconv & 256X256X128 & 64 & 2/3X3 & 512X512X64 \\
\hline
28 & Decode & concat & 512X512X64 &  &  & 512X512X128 \\
\hline
29 & Decode & conv & 512X512X128 & 64 & 1/3X3 & 512X512X64 \\
\hline
30 & Decode & conv & 512X512X64 & 64 & 1/3X3 & 512X512X64 \\
\hline
31 & Decode & conv & 512X512X64 & 3 & 1/1X1 & 512X512X3 \\
\hline
\end{tabular}
\end{table}

\subsection{Loss function}

Dice Loss has been proven as a suitable loss function for medical image segmentation. Dice Loss function is defined in eq(\ref{dice_loss}):
\begin{equation}
{
DL(P,P^\prime)=1-\frac{2 |P\cap P^\prime |}{|P|+|P^\prime|}
}
\label{dice_loss}
\end{equation}
where P represents the predicted result, and $P^\prime$ represents the ground truth annotation. However, in our case, the proportion of calcified voxels among overall annotation voxels is very low, this could introduce class unbalance in the training process. Class unbalance can cause biased predictions, which may result in high specificity but low sensitivity. Prediction bias is unacceptable, especially in medical applications where false positives are much more tolerable compared to false negatives. Several methods have been proposed to deal with this problem including balanced sampling, two-step training, sample re-weighting, and similarity loss functions. To address the class imbalance effect \cite{zhang2021overview}, we have also Tversky loss  function \cite{kamal2019automatic} as defined in eq(\ref{tversky_loss})
 
\begin{equation}
{
TL(P,P^\prime)=1-\frac{|P\cap P^\prime |}{|P\cap P^\prime | + \alpha|P-P^\prime|+\beta|P^\prime - P|}
}
\label{tversky_loss}
\end{equation}
where $\alpha$ and $\beta$ are the hyperparameters that are used to tune false negative and false positive toleration rates, which must satisfy $\alpha + \beta = 1$. When $\alpha = \beta = 0.5$, Tversky loss is the same as the Dice loss

\section{Comparative analysis}
\label{sec.4}

\subsection{Dataset description}

We used CBIS-DDSM Breast Cancer Image Dataset \cite{lee2017curated}, 

The dataset comprises 2,620 digitally scanned film mammography studies, among them 753 calcification cases. CBIS-DDSM encompasses verified pathology information across normal, benign, and malignant cases. The extensive size of the database, coupled with its validated ground truth, renders it a valuable asset for the advancement and evaluation of decision support systems.
As IP-UNet and 3D-UNet are supposed to be trained using 3D volumes, we have converted the CT scans to NIfTI data format \cite{nifti} using dicom2nifti python library \footnote{www.github.com/icometrix/dicom2nifti}. 

\begin{figure}[!ht]
	\centering
	\includegraphics[scale=1]{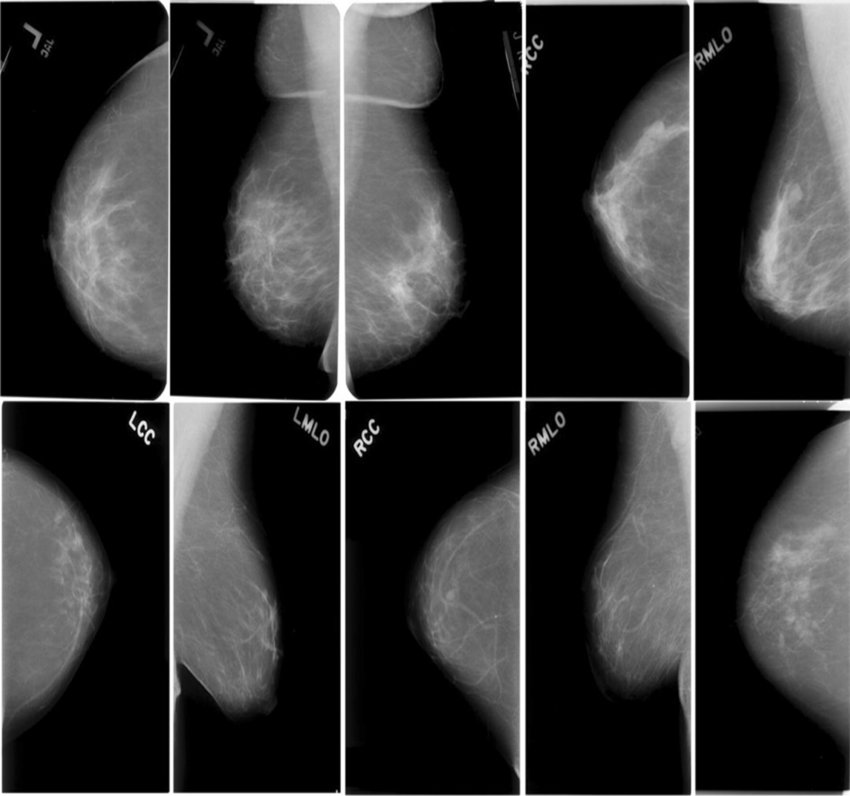}
	\caption{Example of scans from CBIS-DDSM dataset  }
	\label{patient_example}
\end{figure}

Figure \ref{patient_example} shows an example of a CT scan displayed in ITK-SNAP. The dataset statistics are shown in Table \ref{dataset_stat}.

\begin{table}
\centering
\caption{Dataset statistics}
\label{dataset_stat}
\begin{tabular}{|l|l|}
\hline
\textbf{Parameter}  & \textbf{Value}\\ \hline
Number of patients  & 753\\ \hline
Benign cases (Training)        & 329 \\ \hline
Malignant Cases (Training)     & 273 \\ \hline
Benign cases (Testing)        & 85 \\ \hline
Malignant Cases (Testing)     & 66 \\ \hline
\end{tabular}
\end{table}

\subsection{Experiment settings}
\subsection{Result analysis}
\begin{figure*}[!htb]
	\centering
	\includegraphics[scale=0.9]{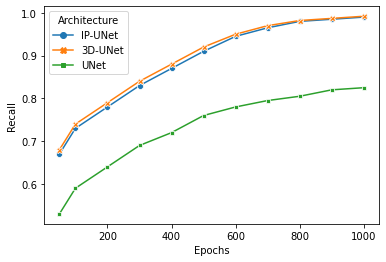}
	\caption{Recall comparison results}
	\label{result_recall}
\end{figure*}
\begin{figure*}[!htb]
	\centering
	\includegraphics[scale=0.9]{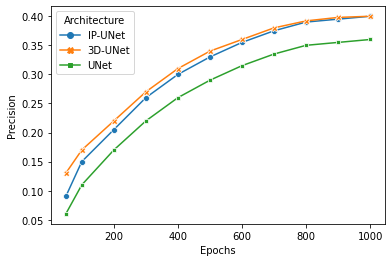}
	\caption{Recall comparison results}
	\label{result_precision}
\end{figure*}

\begin{figure*}[!htb]
	\centering
	\includegraphics[scale=0.9]{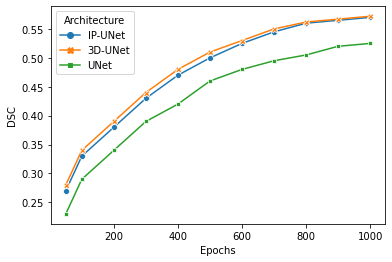}
	\caption{DSC comparison results}
	\label{result_dice}
\end{figure*}

As mentioned earlier, we have compared IP-UNet with the conventional UNet, and a 3D-UNet. The architectures of the baselines have been set as follows:
(1) UNet: Each DICOM slice is inputted into 2D UNet as defined in Table \ref{unet_parameters}. The segmentation output of all DICOM series is merged into a single 3D image and compared with the ground-truth annotation to compute the Dice score similarity. 
(2) 3D-UNet: Implemented 3D-UNet model that takes the NIfTi 3D as input (nii file), or an aggregated dicom series as input. The patch size is 512X512X128, and the initial learning rate was $10^{-3}$.
IP-UNet and both baselines have been trained for 1000 epochs. All experiments were conducted with Alienware workstation with Intel(R) Core(TM) i7-8700K CPU @ 3.70GHz, equipped with 2 GPUs (GTX 1080 Ti).

\subsection{Evaluation metrics}

We used Precision, Recall and Dice Similarity Coefficient (DSC) as the evaluation metrics.

\textbf{Precision:} also known as sensitivity, and it refers to the proportion of positives that are correctly identified compared to the ground truth, and is computed as shown in (\ref{precision_eq}), where $TP$ is the number of true positives and $FN$ is the number of false negatives.

\begin{equation}
{Precision=\frac{TP}{TP+FN}}
\label{precision_eq}
\end{equation}

\textbf{Recall:} also known as specificity, and it measures the proportion of negatives that were correctly predicted, and is computed as shown in (\ref{recall_eq}), where $TN$ is the number of true negatives and $FP$ is the number of false positives.

\begin{equation}
  {recall =\frac{TN}{T N+FP}}
  \label{recall_eq}
\end{equation}

\textbf{DSC:} is used to quantify the similarity between the predicted segmentation and the ground truth annotations, and is computed as shown in (\ref{dsc_eq}), where 
where $TP$ is the number of true positives, $FN$ is the number of false negatives 

\begin{equation}
{
DSC=\frac{2 |P\cap P^\prime |}{|P|+|P^\prime|} = \frac{2TP}{2TP+FN+FP}
}
\label{dsc_eq}
\end{equation}

\begin{figure*}[!ht]
	\centering
	\includegraphics[scale=0.9]{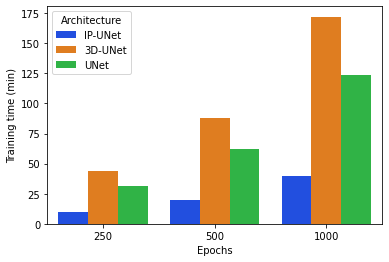}
	\caption{Training time comparison}
	\label{result_training}
\end{figure*}

\begin{figure*}[!ht]
	\centering
	\includegraphics[scale=0.9]{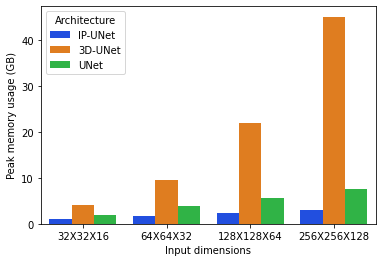}
	\caption{Memory requirement comparison}
	\label{result_memory}
\end{figure*}

Figure \ref{result_recall}, Figure \ref{result_precision} and Figure \ref{result_dice} compare IP-UNet to other baselines in terms of recall, precision and DSC respectively. In Figure \ref{result_recall}, all architectures have relatively low recall when trained for less than 500 epochs. After that they significantly improve, to finally stabilize after 750 epochs. 3D-UNet and IP-UNet have much higher recall compared to the conventional 2D UNet. 3D-UNet and IP-UNet score recall of $99\%$, while 2D UNet scores as low as $82\%$. Such high recall achieved by 3D-UNet and IP-UNet is misleading, as precision does not take FN into account, and in our case, the proportion of calcified voxels among overall annotation voxels is very low, this results in a very high recall but low precision.

Figure \ref{result_precision} shows the precision of IP-UNet compared to the studied baselines. Again, 3D-UNet and IP-UNet have much higher precision compared to the conventional 2D UNet. But 3D-UNet and IP-UNet score just about $40\%$ due to the high FN cases, especially for low calcification patients. Finally, Figure \ref{result_dice} shows the DSC for UNet, 3D-UNet and IP-UNet, the latter scores $57\%$ when trained for more than 750 epochs. Table \ref{results_table} shows the final recall, precision and DSC of the studied architecture when trained for 1000 epochs.

\begin{table}[]
\centering
\caption{Sensitivity, specificity and Dice similarity coefficient}
\begin{tabular}{|l|l|l|l|}
\hline
\textbf{System} & \textbf{Recall} & \textbf{Precision} & \textbf{DSC} \\ \hline
IP-UNet & 0.99   & 0.4 & 0.525  \\ \hline
UNet & 0.825 & 0.36 & 0.525     \\ \hline
3D-UNet & 0.992 & 0.4 & 0.572   \\ \hline
\end{tabular}
\label{results_table}
\end{table}

Although our proposed architecture IP-UNet has similar recall, precision and DSC with 3D-UNet, nonetheless, it is much more computationally efficient as it is trained with intensity projections rather than the full 3D volumic data. Figure \ref{result_training} shows the required training time for IP-UNet, 3D-UNet and 2D UNet respectively. IP-UNet maintains low training time, even when the model is trained with 1000 epochs. Unlike 3D-UNet and 2D UNet where the training time exponentially increases in proportion to the increase of training epochs. Similarly, Figure \ref{result_memory} shows the average memory consumption of the studied models. Memory consumption was measured using tracemalloc Python package, and the values presented in Figure \ref{result_memory} are the average of 10 training cycles.

\section{Conclusion}
\label{sec.5}
In this paper, we have proposed IP-UNet, an end-to-end deep learning model that operates on intensity projections of 3D volumes instead of memory-consuming 3D volumetric data. IP-UNet can be trained with limited memory capability without loss of resolution of the original 3D images. We compared the performance of IP-UNet in terms of segmentation accuracy and computational cost with similar baselines, namely, UNet and 3D-UNet. In the experiment, we trained and evaluated IP-UNet on 500 3D volumetric images for automatic breast calcification detection. Results show that IP-Unet can achieve similar segmentation accuracy with 3D-Unet, and reduces the training time by 70\%, and memory requirement by 92\%.


\section*{Acknowledgment}
This work was supported by:
The Insight Centre for Data Analytics funded by Science Foundation Ireland under Grant Number 12/RC/2289 P2.



\bibliographystyle{IEEEtran}
\bibliography{refs}
%

\vskip 0pt plus -1fil
\begin{IEEEbiography}[{\includegraphics[width=1in,height=1.25in,clip,keepaspectratio]{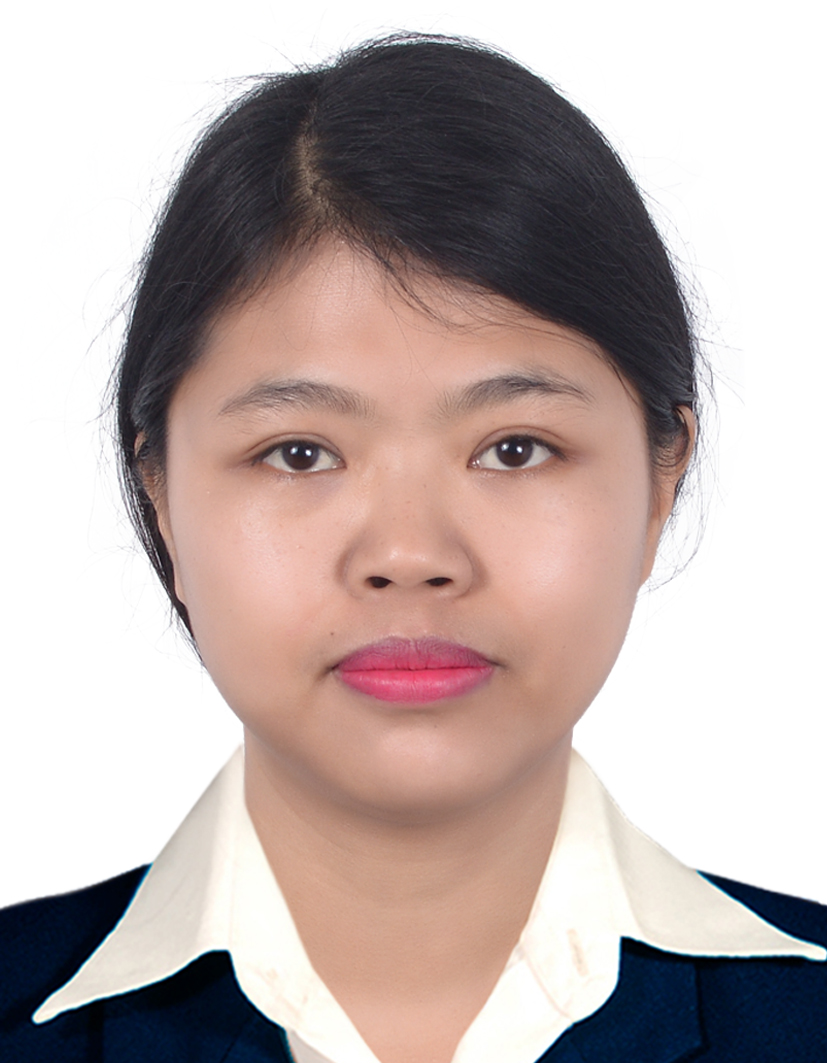}}]{Nyothiri Aung}
is a postdoctoral researcher at University College Dublin, Ireland. She received her PhD in Computer Science and Technology from University of Science and Technology Beijing, China, 2020. And a Master's of Information Technology from Mandalay Technological University, Myanmar, 2012. She worked as a demonstrator at the Department of Information Technology in Technological University of Meiktila, Myanmar (2008-2010). Her research interests include Social Computing, Medical image analysis, and Intelligent Transportation Systems.
\end{IEEEbiography}

\vskip 0pt plus -1fil
\begin{IEEEbiography}[{\includegraphics[width=1in,height=1.25in,clip,keepaspectratio]{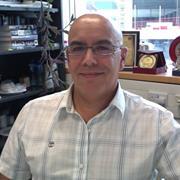}}]{Mohand Tahar Kechadi}
is a full professor in school of computer science, University College Dublin, Ireland. He received master’s and Ph.D. degrees in computer science from the University of Lille 1, France. His research interests include data mining, distributed data mining heterogeneous distributed systems, grid and cloud computing, and digital forensics and cyber-crime investigations. He is a member of the Communications of the ACM journal and IEEE Computer Society. He is an Editorial Board Member of journal of Future Generation Computer Systems.
\end{IEEEbiography}

\vskip 0pt plus -1fil

\begin{IEEEbiography}[{\includegraphics[width=1in,height=1.25in,clip,keepaspectratio]{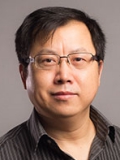}}]{Liming Chen}
is a professor in the School of Computer  Science  and  Informatics  at  University  of  Ulster,  Newtownabbey,  United  Kingdom.  He  received his  B.Eng  and  M.Eng  from  Beijing  Institute  of Technology  (BIT),  Beijing,  China,  and  his  Ph.D  in Artificial Intelligence from De Montfort University,UK.  His  research  interests  include  data  analysis,ubiquitous computing, and human-computer interaction. Liming is a Fellow of IET, a Senior Member of IEEE, a Member of the IEEE Computational Intelligence Society (IEEE CIS), a Member of the IEEE CIS Smart World Technical Committee (SWTC), and the Founding Chair of the IEEE CIS SWTC Task Force on User-centred Smart Systems (TF-UCSS). He has served as an expert assessor, panel member and evaluator for UK EPSRC (Engineering and Physical Sciences Research Council, member of the Peer Review College), ESRC (Economic and Social Science Research Council), European Commission Horizon 2020 Research Program, Danish Agency for Science and Higher Education, Denmark, Canada Foundation for Innovation (CFI), Canada, Chilean National Science and Technology Commission (CONICYT), Chile, and NWO (The Netherlands Organisation for Scientific Research), Netherlands.
\end{IEEEbiography}

\vskip 0pt plus -1fil
\begin{IEEEbiography}[{\includegraphics[width=1in,height=1.25in,clip,keepaspectratio]{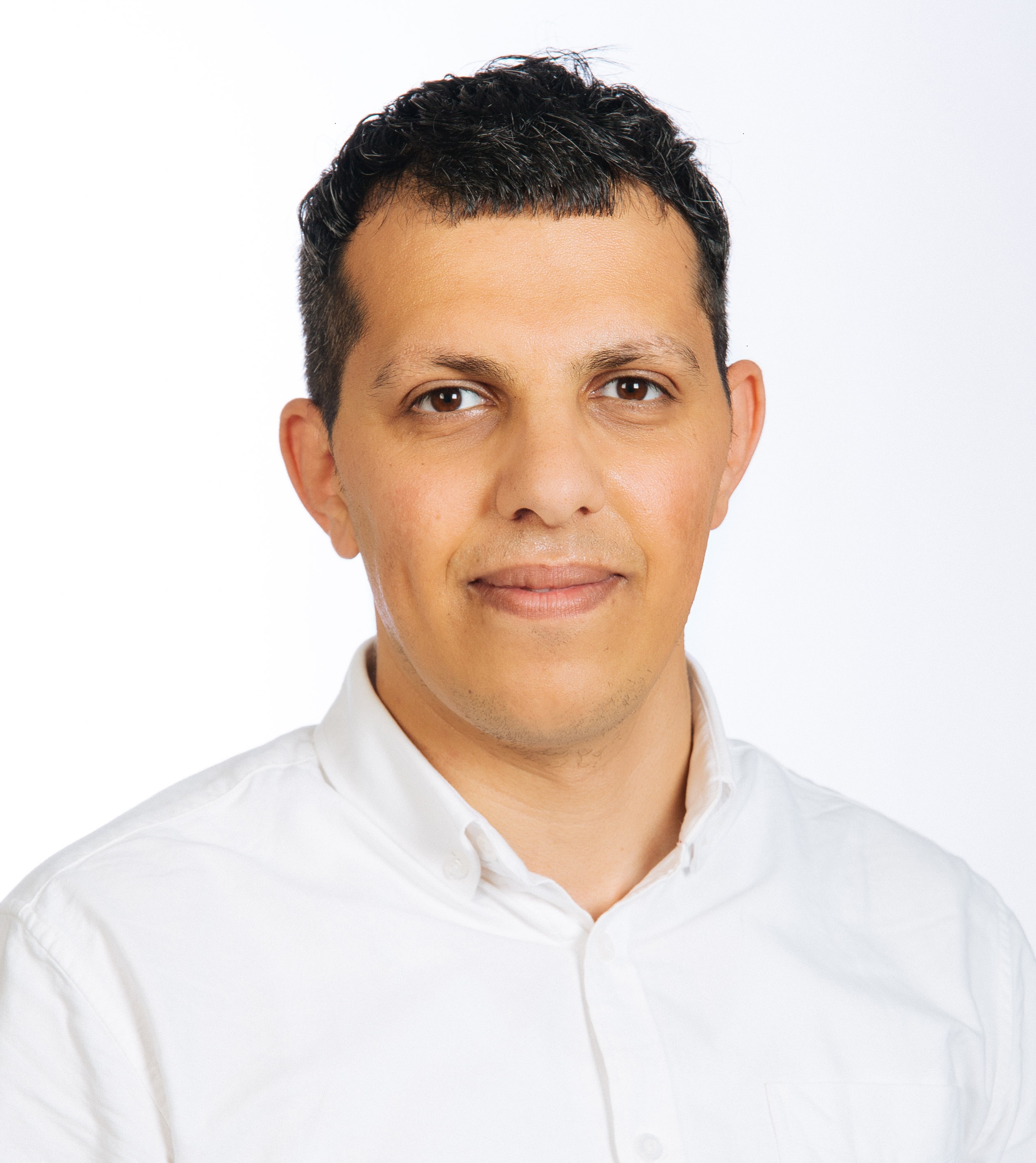}}]{Sahraoui Dhelim} is a senior postdoctoral researcher at University College Dublin, Ireland. He was a visiting researcher at Ulster University, UK (2020-2021). He obtained his PhD degree in Computer Science and Technology from the University of Science and Technology Beijing, China, in 2020. And a Master's degree in Networking and Distributed Systems from the University of Laghouat, Algeria, in 2014. And Bs degree in computer science from the University of Djelfa, in 2012. He is a guest editor in several reputable journals, including Electronics Journal and Applied Sciences Journal. His research interests include Social Computing, Smart Agriculture, Deep-learning, Recommendation Systems and Intelligent Transportation Systems.
\end{IEEEbiography}
\end{document}